\documentclass[12pt]{article}
\usepackage{amsmath}
\usepackage{epsfig}

\begin{document}
\begin{titlepage}
\begin{center}
{\large \bf  Simple models of the contribution of
intermediate stage gluons to $J/\psi$ suppression
at the CERN SPS}
\end{center}

\begin{center}
{M.Jur\v{c}ovi\v{c}ov\'a${}^{a)}$, A.Nogov\'a${}^{b)}$,
J. Pi\v{s}\'ut${}^{a)}$,

 N. Pi\v{s}\'utov\'a${}^{a)}$, and K. Tok\'ar${}^{a)}$ }
\end{center}

\begin{center}
${}^{a)}${\it Department of Physics, Comenius University,

 SK-84248 Bratislava, Slovakia}
\end{center}
\begin{center}
${}^{b)}${\it Institute of Physics of the Slovak Academy of Science,

 SK-84228 Bratislava, Slovakia}
\end{center}

\abstract%
{ We construct three simple scenarios of the
time -- dependence of
density of intermediate stage gluons in nuclear
collisions in the CERN SPS energy range. Gluons with
 energy of about 0.6 -- 1.0 GeV are assumed to be
produced in nucleon--nucleon collisions in a Glauber
type model. The rate of gluon production is given by the
parameter $n_{g/nn}$ equal to the average number of gluons
produced per nucleon - nucleon collision. The value of this
parameter determines the behaviour of the gas of gluons.
The number of gluons increases due to gluon branching
and processes like g+g$\rightarrow $g+g+g and decreases  
due to the hadronization. Gluons are assumed to be able to
dissociate J/$\psi$ in g+J/$\psi$ collisions, the
dissociation cross--section $\sigma_{g\psi}$ is taken as
a free parameter. In the first scenario, the energy
density of the gas of gluons never reaches the critical
energy density $\epsilon_c \approx 0.7$ GeV/fm$^3$ and
gluons rapidly hadronize. In the second scenario, the
critical energy density is reached but the system of gluons
is unable to reach thermalization.
In the third scenario gluons reach thermalization and the
thermalized system suppresses J/$\psi$ by the Matsui--Satz
mechanism.
The third scenario under the assumption of a small value of
$\sigma_{g\psi}$ is able to describe qualitatively the data on
$J/\psi$ suppression in
Pb--Pb interactions obtained by the NA50 Collaboration.
Other scenarios have problems with getting the rather
abrupt onset of $J/\psi$ suppression.}
\end{titlepage}

\section{Introduction}
\label{intro}

    Suppression of $J/\psi$ in heavy-ion collisions
has been suggested by
Matsui and Satz \cite{MS86} as a signature
of the Quark-Gluon Plasma (QGP)
formation in heavy ion collisions. The observation
 of $J/\psi$ suppression
by NA38 and NA50 collaborations \cite{NA50a,NA50b}
in collisions of lighter beams,
up to ${}^{32}S$, with heavy targets has been basically
explained by Gerschel and H\"ufner, and Capella et al.
\cite{GH88,GH92,GH99} as due to the
disintegration of $J/\psi$ by nucleons present
in the colliding ions. We shall refer to this mechanism as to the
nuclear absorption (NA).
Some role  might also be
played  by the disintegration of $J/\psi$ by collisions with
secondary hadrons produced during the collisions
\cite{BM88,FLP88,VG,SVG99,ACF99,RV}.
     The "anomalous" $J/\psi$ suppression observed by the
NA50 collaboration in Pb-Pb collisions at the CERN-SPS at
 $E_{Lab}$= 160GeV/nucleon, cannot be
explained by nuclear absorption alone and models based
on $J/\psi$
disintegration by secondary hadrons meet difficulties \cite{RV}.

     Simple models of anomalous $J/\psi$ suppression
 by QGP have been proposed
by Blaizot and Ollitrault \cite{BO96,BO2000}
(in what follows referred to as the
BO model) and by Kharzeev, Louren\c{c}o, Nardi and
 Satz \cite{KLNS97,Satz} (KLNS model).
Both models are based on the dynamics of tube-on-tube collisions
 within the Glauber model and
assume that in a given tube-on-tube collision QGP
is produced provided that a certain condition involving the
lengths of the two tubes is satisfied.

A combination of $J/\psi$ suppression by QGP, by
 nuclear absorption and by
hadron gas has been considered by Chaudhuri \cite{Chaud}.
Kharzeev and Satz have argued \cite{KSetal} that
cross--section for disintegration of $J/\psi$ by
thermal hadrons are very low and that the contribution
 to $J/\psi$ suppression by this mechanism is negligible.
The arguments seem to be convincing but the question is
 not yet solved
completely.

Another possible source of anomalous $J/\psi$
 suppression may be due to
$J/\psi$ disintergration by the intermediate stage
gluons. The evolution of gluon cascades  has been studied e.g. in
Monte Carlo cascade codes \cite{G95,GS97,Bass99}.

     The purpose of the present paper is to study $J/\psi$
suppression by Intermediate Stage Gluons (ISG).  Model
of this type has been discussed
in Ref.\cite{FPP} but it was oversimplified
and no definite conclusions could have been reached.

 We shall consider here three simple scenarios of $J/\psi$
suppression in Pb--Pb collisions at the CERN SPS. In each of them
$J/\psi$ is suppressed by the gluon gas and by nuclear
absorption. Only in the third one there is the additional
suppression by the thermalized system of gluons, described
by the Matsui -- Satz mechanism.

(i) In the first scenario, the energy
density of the gas of gluons never reaches the critical
energy density $\epsilon_c \approx 0.7$ GeV/fm$^3$ and
gluons rapidly hadronize.

(ii)In the second scenario, the
critical energy density is reached but the system of gluons
is unable to reach thermalization, since the energy density
becomes subcritical before the system is thermalized.

(iii) In the third scenario gluons reach thermalization
in collisions of tubes of appropriate lengths  and the
thermalized system suppresses J/$\psi$ by the Matsui--Satz
\cite{MS86} mechanism.
Under the assumption of a small value of
$\sigma_{g\psi}$ this scenario is able to describe the data on
$J/\psi$ suppression in
Pb--Pb interactions obtained by the NA50 Collaboration.

Anticipating the results described
below we find it  rather improbable that
 $J/\psi$ disintegration by intermediate stage gluons
provides a substantial part of  $J/\psi$ suppression
as observed by the NA50 collaboration in Pb--Pb interactions.
In this sense our results give an indirect support
 to the idea that a phase
transition to QGP is responsible for the anomalous $J/\psi$
 suppression. The key arguments is
 that the onset of anomalous $J/\psi$ suppression
 is too "abrupt" to be ascribed to the ISG which do not reach
the thermalized state.

     The paper is organised as follows. In Sect.2 we
briefly discuss different contributions to the $J/\psi$ suppression
and give the corresponding formulas.
     In Sect.3 we describe the three scenarios of the behaviour of
intermediate stage gluons (ISG). $J/\psi$ suppression by ISG
in the three scenarios is discussed in Sect. 4.
Reasons why  scenarios without thermalized gluons cannot
describe the NA50 data are quite
simple - the suppression as function of $E_T$ is "too continuous",
 whereas the data
seem to indicate some abrupt increase of $J/\psi$ suppression.
Comments and conclusions are deferred to Sect.5.
Some details concerning possible sources of intermediate
 stage gluons are presented in the Appendix.

\section{ Contributions to $J/\psi$ suppression}
\label{DM}

In a simple model of spherical nuclei with constant nucleon density
the $J/\psi$ survival probability $S^{J/\psi}_{AB}(b)$,
where b is the impact parameter, is given as
\begin{equation}
S^{J/\psi}_{AB}(b)=\frac {\sigma^{J/\psi}_{AB}(b)}{n^{AB}_{coll}(b)\sigma^{J/\psi}_{nn}}
\label{eq1}
\end{equation}
where the total $J/\psi$ production cross-section in $AB$ interaction is denoted
as $\sigma^{J/\psi}_{AB}(b)$ and $\sigma^{J/\psi}_{nn}$ is the
average $J/\psi$ production cross-section in
nucleon-nucleon collision. Frequently discussed
 contributions to the $J/\psi$
survival probability are given by the following expression
\begin{equation}
 S=\frac {N}{N_0}
\label{eq2}
\end{equation}
The numerator $N$ is given as
$$
N=\int_0^{R_A}sds\int_0^{2\pi}d\theta
\int_{-L_A(s)}^{L_A(s)}dz_A\int_{-L_B(s,\theta)}^{L_B(b,s,\theta)}dz_B
$$
$$
e^{-\rho_A\sigma_a[z_A+L_A(s)]}e^{-\rho_B\sigma_a[z_B+L_B(b,s,\theta)]}
$$
\begin{equation}
F_{sh}F_{hJ/\psi}F_{gJ/\psi}F_{QGP}
\label{eq3}
\end{equation}
where
$$
L_A(s)=\sqrt{R^2_A-s^2},\quad
L_B(b,s,\theta)=\sqrt{R_B^2-b^2-s^2+2bscos(\theta)}
$$
when the expression under the square-root in $L_B$ is positive, and
$L_B(b,s,\theta)=0$ when this expression is negative.

The term containing $\sigma_a$ corresponds to the nuclear absorption
 with $\sigma_a$ being the absorption
cross--section and the terms $F_{sh},
F_{hJ/\psi},F_{gJ/\psi},F_{QGP}$ correspond to the shadowing of
gluon structure functions, $J/\psi$ suppression by secondary hadrons,
$J/\psi$ suppression by intermediate stage gluons and
$J/\psi$ suppression by QGP, respectively.
$N_0$ is obtained from $N$ by putting $\sigma_a=0$ and
$F_{sh}=F_{hJ/\psi}=F_{gJ/\psi}=F_{QGP}=1$.

The geometrical notation used in Eq.(3) is standard,
 $\vec s$  is the vector connecting
the position in the transverse
plane to the center of nucleus  $A$  and $\theta$
is the angle between $\vec b$ and $\vec s$.

In what follows we shall consider the $AB$ interaction
in the center of mass of nucleon-nucleon
collisions, thus $L_A$ and $L_B$ will be contracted by
the Lorentz factor $\gamma$ and nuclear densities $\rho_A$
and $\rho_B$ will be multiplied by $\gamma$. We shall use
the notation $l_A = L_A/\gamma$ and $l_B=L_B/\gamma$

The position of a nucleon within the tube in nucleus $A$ is denoted
as $z_A$, $-l_A\le z_A\le l_A$, in nucleus $B$ as $z_B$,
$-l_B\le z_B\le l_B$
with $z_A,z_B$ increasing in the direction
 of motion of nuclei $A,B$.
Nucleon densities $\rho_A,\rho_B$  are given
as $\rho_A=\gamma \rho_0$,
$\rho_B=\gamma \rho_0$, where $\rho_0$
is the density of nucleons in a nucleus at rest,
$\rho_0 = 0.138 fm^{-3}$,
corresponding to a spherical nucleus with the radius
$R_A=1.2A^{1/3}fm$.

Comparison with earlier as well as with recent
NA50 data \cite{NA50Nantes} indicates that nuclear absorption (NA)
can describe $J/\psi$ suppression
up to S+U interactions and only Pb-Pb case cannot
be described by NA.
     According to the Blaizot-Ollitrault (BO) \cite{BO96}
mechanism of $J/\psi$ suppression by QGP,
all $J/\psi$'s produced in
tube-on-tube collisions are
totally absorbed, provided that
\begin{equation}
\kappa_{BO}(b,s,\theta)=\rho_A\sigma_{nn}2l_A(s)+\rho_B\sigma_{nn}
2l_B(b,s,\theta)\ge \kappa_{BO,c}
\label{eq4}
\end{equation}
where $\kappa_{BO,c}=9.75$ and $\sigma_{nn}$ denotes the
non--difractive nucleon--nucleon cross--section.
   We are using here the notation of
 Ref.\cite{NPP}. In this case, the factor $F_{QGP}$ in Eq.(3)
becomes simply
\begin{equation}
F_{QGP}=\Theta(\kappa_{BO,c}-\kappa_{BO}(b,s,\theta))
\label{eq5}
\end{equation}
where $\Theta(x)$ is the Heavyside function.

In the Kharzeev, Nardi, Louren\c{c}o, and Satz (KLNS)
 \cite{KLNS97} model
of $J/\psi$ suppression by QGP
 one defines
\begin{equation}
\kappa(b,s,\theta)_{KS}=\frac
{\rho_A\sigma_{nn}2l_A(s).\rho_B\sigma_{nn}2l_B(b,s,\theta)}
{\rho_A\sigma_{nn}2l_A(s)+\rho_B\sigma_{nn}2l_B(b,s,\theta)}
\label{eq6}
\end{equation}
and it is
assumed  that all $J/\psi$'s produced in
a tube-on-tube interaction
are completely suppressed, provided that
\begin{equation}
\kappa_{KS}(b,s,\theta)\ge \kappa_{KS,c}
\label{eq7}
\end{equation}
This is equivalent to the factor
\begin{equation}
F_{QGP}=\Theta(\kappa_{KS,c}-\kappa_{KS}(b,s,\theta))
\label{eq8}
\end{equation}
in Eq.(3).  In their model KLNS
take into account that $60\%$ of $J/\psi$'s in the final state
 is produced directly (after $\psi$' decays
have been subtracted, since $\psi$' is suppressed
already in S+U interactions)
and $40\%$ come from $\chi_c$ decays. The value
$\kappa_{KS,c}= 2.43$
 corresponds  to $\chi_c$
suppression and the threshold for the supression of
directly produced $J/\psi$ is higher. In our notation
the second threshold would appear at about
$\kappa_{KS,c}^{J/\psi}\approx 2.9$. The corresponding $F_{QGP}$
in Eq.(3) would then become
\begin{equation}
F_{QGP}=0.4\Theta(2.43-\kappa_{KS}(b,s,\theta))+
0.6\Theta(2.9-\kappa_{KS}(b,s,\theta))
\label{eq9}
\end{equation}
Since we are mostly interested in the first threshold
we shall use below $F_{QGP}$
as given by Eq.(8). The value of the transverse energy
 $E_T$ (in GeV) in NA50 experiments is approximately given as
\begin{equation}
E_T(b)=0.325N_w(b)
\label{eq10}
\end{equation}
where $N_w(b)$ is the number of participating
(wounded) nucleous in a Pb-Pb collision at the
impact parameter $b$.

The onset of anomalous $J/\psi$ suppression as given by both
BO and KLNS models is rather abrupt and it seems that this
is also required by the data, see, however, Refs.\cite{QIU,AKC}.

\section{Model of the space and time
dependence of density of intermediate stage gluons}
\label{model}

Intermediate stage gluons are supposed to be produced
in individual nucleon-nucleon
collisions by the mechanism $g+g\rightarrow g+g$.
The arguments supporting this possibility are
summarized in Appendix A. We assume that the dynamics
of the formation of the gluon gas
is --at least in the first stage of the collision when nuclei
 pas through
each other -- basically longitudinal and that we can treat
separately gluons produced in different
tube-on-tube collisions. Some support to this simplification
comes from the fact that
 $\sigma(g+g\to g+g)$ is peaked in forward and backward directions.
In the calculations
below we shall use the c.m.s. of nucleon-nucleon collision. The
tubes pass through each other at the time
$$t_1=l_A+l_B$$
 The total number of
nucleon-nucleon collisions before time $t_1$ is
\begin{equation}
N_{nn}=\rho_A\sigma_{nn}2l_A(s).\rho_B\sigma_{nn}2l_B(b,s,\theta)
\label{eq11}
\end{equation}
where $\rho_A=\rho_B=\gamma \rho_0$ and $\rho_0$ is the nucleon
density for nucleus at rest. At the time $t_1$ the volume of the two tubes is
$V = (2l_A+2l_B)\sigma_{nn}$. Denoting the
average number of gluons per nucleon-nucleon collision as $n_{g/nn}$
and assuming the
density of gluons to be homogeneous within the two tubes, just after
they have passed through each other,
 we obtain for the density of gluons at the time $t_1$
\begin{equation}
n(t_1)=\frac{\rho_A\sigma_{nn}2l_A(s).
             \rho_B\sigma_{nn}2l_B(b,s,\theta)}
            {(2l_A+2l_B)\sigma_{nn}}
            n_{g/nn}
\label{eq12}
\end{equation}
For $t\le t_1$ we shall assume that
\begin{equation}
n(t)={\frac {t}{t_1}}n(t_1), \qquad t\le t_1
\label{eq13}
\end{equation}
Gluons, when produced in nucleon--nucleon collisions
are supposed to have energies mostly in the interval
of 0.6 -- 1.0 GeV and for estimates we shall take the
average energy of a gluon at $t_1$ as $E_g(t_1)$ = 0.8 GeV.
The energy density of the system of gluons at the time $t_1$
corresponding to the density in Eq.(12) then becomes
\begin{equation}
 \epsilon_g(t_1) = 0.8 GeV.n(t_1)
\label{eq14}
\end{equation}
Depending on the value of $\epsilon(t_1)$ we shall define below
three regimes

i) The energy density $\epsilon(t_1)$ is lower than
the critical value $\epsilon(t_1)\le \epsilon_{c}$,
where $\epsilon_{c}$ corresponds to the energy density
of the phase transition to thermalized gluons or QGP.
For numerical estimates we shall take here the value
obtained in the lattice calculations $\epsilon_{c}$ =
0.7 GeV/fm$^3$.
In this case the system of gluons rapidly hadronizes. We shall
describe the hadronization by
\begin{equation}
n(t)=n(t_1)\exp\left(-\frac{t-t_1}{\tau_h}\right)
\label{eq15}
\end{equation}
where $\tau_h$ is the hadronization time. For numerical
estimates we shall take $\tau_h$ = 1fm/c.

ii) The energy density $\epsilon(t_1)$ is larger than $\epsilon_{c}$,
but due to the longitudinal expansion the energy density of ISG becomes lower than $\epsilon_{c}$ before ISG thermalizes.
Supposing that ISG expands with velocity $v_0$ (we have in mind
the rapidity interval $-0.5 \le y \le 0.5$ and $v_0/2$ is the velocity
at y=0.5 ). In the c.m. frame of nucleon--nucleon collision we have
from the energy conservation
\begin{equation}
\epsilon(t)=\epsilon (t_1)\frac {2l_A+2l_B}{2l_A+2l_B+v_0(t-t_1)},
\quad t\ge t_1
\label{eq16}
\end{equation}
denoting the thermalization time as $t_{Th}$, the condition for
reaching the thermalization becomes
\begin{equation}
\epsilon(t_1+t_{Th})\ge \epsilon_{c}
\label{eq17}
\end{equation}
If this is not true the system will reach the energy density
$\epsilon_{c}$ before it is thermalized. In that case there exists
the time $t_2\le t_1+t_{Th}$ for which it holds
\begin{equation}
\epsilon(t_2)=\epsilon_{c}= \epsilon(t_1)
\frac{2l_A+2l_B}{2l_A+2l_B+v_0(t_2-t_1)}
\label{eq18}
\end{equation}
From Eq.(18) we find
\begin{equation}
t_2=t_1+{\frac{1}{v_0}}
\left[\frac{\epsilon(t_1)}{\epsilon_{c}}-1\right]
(2l_A+2l_B)
\label{eq19}
\end{equation}
The system thus does not reach thermalization provided that
\begin{equation}
{\frac{1}{v_0}}\left[\frac{\epsilon(t_1)}{\epsilon_{c}}-1\right]
(2l_A+2l_B)\le t_{Th}
\label{eq20}
\end{equation}
During the approach to thermalization the average energy per gluon
will decrease from the original value of about 0.8 GeV to the
thermal value of about 0.2 GeV. This can happen by the branching of
originally produced off--the--mass--shell gluons or by processes
like g+g$\rightarrow$ g+g+g. Since the emission of a gluon with
energy of about $\Delta E\sim $ 0.2 GeV takes the time interval $\Delta t$ $\sim$
$\hbar/\Delta E \sim$ 1fm/c we expect that the time of thermalization
$t_{Th}$ satisfies the condition
\begin{equation}
2fm/c \le t_{Th} \le 3fm/c
\label{eq21}
\end{equation}
for numerical estimates we shall take $t_{Th}$ = 2.6 fm/c.
In the case that $t_2-t_1 \le t_{Th}$ the density of gluons within
the time interval $t_1 \le t \le t_2$ will be described as
\begin{equation}
n(t)= \left[n(t_1)+{\frac{t-t_1}{\tau_{em}}}n(t_1)\right]
\frac{2l_A+2l_B}{2l_A+2l_B+v_0(t-t_1)}
\label{eq22}
\end{equation}
and at $t_2$ given by Eq.(19) the density of gluons becomes
\begin{equation}
n(t_2)= \left[n(t_1)+{\frac{t_2-t_1}{\tau_{em}}}n(t_1)\right]
\frac{2l_A+2l_B}{2l_A+2l_B+v_0(t_2-t_1)}
\label{eq23}
\end{equation}
where $\tau_{em}\approx$ 1 fm/c is the time for the emission of
an additional  gluon, we shall take $\tau_{em}=$ 1 fm/c.
Since thermalization has not been reached the system after
the time $t_2$ hadronizes according to
\begin{equation}
n(t)= n(t_2)\exp \left(-\frac {t-t_2}{\tau_h}\right)
\label{eq24}
\end{equation}

(iii) In this case ISG reaches thermal equilibrium.
For $t_1\le t \le t_1+t_{Th}$ the density of gluons depends on time
according to Eq.(22). We shall assume that in this case $J/\psi$
will be suppressed by the Matsui--Satz mechanism \cite{MS86}. Since
the mechanism is supposed to dissolve $J/\psi$ rapidly, we do not need
to specify further evolution of n(t) for these tube--on--tube collisions.

The density of gluons corresponding to the three possible types of the
evolution of gluon densities is shown in Fig.1, where we considere
collisions of the longest possible tubes in central Pb--Pb interaction.

Note that the time of thermalization in the case (a) in Fig. 1 is about
4 fm/c what is probably the maximum one can expect to be reasonable for
a purely longitudinal expansion of the system.
\begin{figure}[t]
\begin{center}
   \epsfig{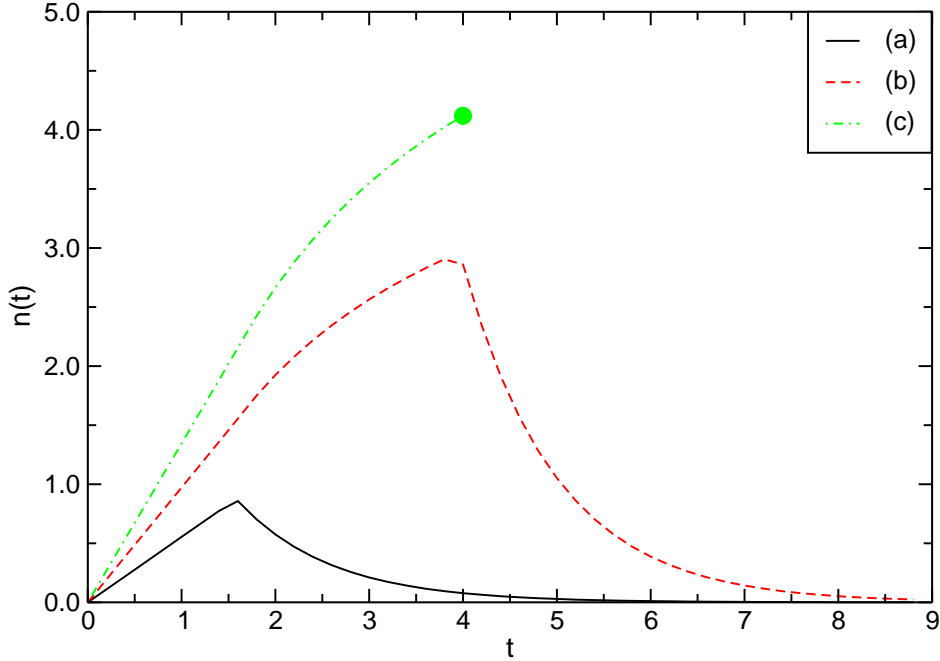}
\end{center}
\caption{Time dependence of the density of gluons n(t)
 in collision of the two
tubes of lengths 2$l_A$ = 2$l_B$ = 2$R_{Pb}/\gamma$ = 1.58 fm.  In all
cases we have used the values of parameters $\epsilon_g$ = 0.8 GeV,
$\tau_{em}$ = 1 fm/c, $\epsilon_{c}$ = 0.7 GeV/fm$^3$. The curve (a)
corresponds to $n_{g/nn}$ = 0.24 (subcritical case),
 (b) refers to $n_{g/nn}$ = 0.42 (supercritical but non-thermalized, and
(c) to $n_{g/nn}$ = 0.58 (thermalized). The time when thermalization
has been reached is denoted by the dot.}
\label{fig1}
\end{figure}

\section {$J/\psi$ suppression by intermediate stage
gluons}
\label{results}

We shall now study $J/\psi$ suppression by intermediate state
gluons. To see the gluon contribution better we shall switch
of all the other contributions.

The suppression factor $F_{gJ/\psi}$
is given as
\begin{equation}
F_{gJ/\psi}=
exp\big\lgroup -\int_{t_0}^\infty v_gn(t)\sigma_{gJ/\psi}dt\big\rgroup
\label{eq25}
\end{equation}
where $t_0$ is the time when $J/\psi$ has been produced in a collision
of two nucleons,
$t_0$ is given by $z_A$ and $z_B$,
$$
t_0= (l_A-z_A+l_B-z_B)/2
$$
 $v_g$  is the relative velocity
of the gluon and $J/\psi$ (we shall take $v_g=1$ in $c=1$ units)
and $\sigma_{gJ/\psi}$ is the averaged
cross-section for the dissociation of $J/\psi$ in
a collision with a gluon.
This expression is valid for each tube - on - tube collision.
In Fig. 2 we show the $E_T$ dependence of the
$F_{gJ/\psi}(E_T)$ factor for cases corresponding to the
function $n(t)$ presented in Fig.1.
\begin{figure}[t]
\begin{center}
  \epsfig{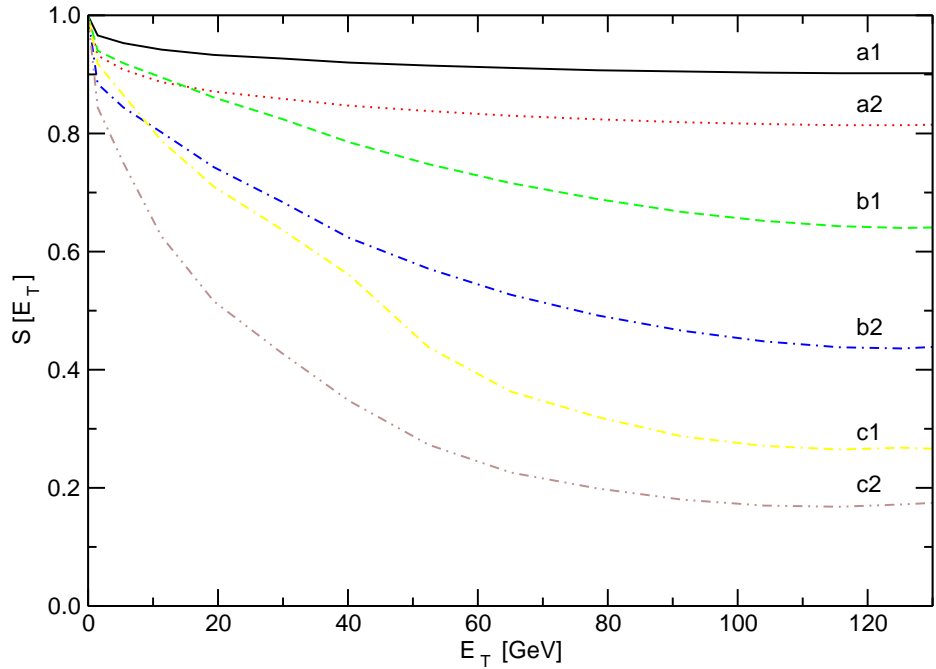}
\end{center}
\caption{The factor $F_{gJ/\psi}(E_T)$ for $J/\psi$ suppression in
Pb--Pb collisions at the SPS. The following  parameters
have the same values in all cases:$\epsilon_g$ = 0.8 GeV,
$\tau_{em}$ = 1 fm/c, $\epsilon_{c}$ = 0.7 GeV/fm$^3$.
Other parameters are different in individual cases. For
(a1) $n_{g/nn}$ = 0.24, $\sigma_{g\psi}$ = 1mb,
(a2) $n_{g/nn}$ = 0.24, $\sigma_{g\psi}$ = 2mb,
(b1) $n_{g/nn}$ = 0.42, $\sigma_{g\psi}$ = 1mb,
(b2) $n_{g/nn}$ = 0.42, $\sigma_{g\psi}$ = 2mb,
(c1) $n_{g/nn}$ = 0.58, $\sigma_{g\psi}$ = 1mb,
(c2) $n_{g/nn}$ = 0.58, $\sigma_{g\psi}$ = 2mb.}
\label{fig2}
\end{figure}
Results in Fig. 2 allow us to make a few conclusions

\begin{itemize}
\item  Cases (a1) and (a2) never lead to energy density larger
than the critical. This results in a rapid hadronization
of gluons. The factor $F_{gJ/\psi}(E_T)$ is rather large
and does not show any rapid onset of $J/\psi$ suppression
at some value of $E_T$. This is true even for rather large
values of the dissociation cross--section $\sigma_{g\psi}$.
\item Cases (b1) and (b2) correspond to larger gluon densities.
Assuming that $\sigma_{g\psi}$ is as large as 1 - 2 mb, the
factor $F_{gJ/\psi}(E_T)$ can go down to 
 values of about 0.5 for large
values of $E_T$. In combination with the effects of nuclear
absorption it could lead to values of $J/\psi$ survival
probability consistent with data. On the other hand this
mechanism does not show any rapid onset of $J/\psi$
suppression at some value of $E_T$
\item In the case of (c1) and (c2) the value of $n_{g/nn}$ is chosen
in such a way that the thermalization of gluon gas appears first
in Pb-Pb collisions at around $E_T$ = 40 GeV. The Matsui -- Satz
mechanism leads to an onset of $J/\psi$ suppression at around
this value of $E_T$. The rapid onset is visible in the curve (c1), but
it is washed out in the curve (c2).
\end{itemize}

We shall now study in more detail the case when dissociation
of $J/\psi$ is present together with the Matsui -- Satz mechanism.
In Fig.3 we present the $J/\psi$ survival probability for
$n_{g/nn}$ = 0.58 and with the nuclear absorption switched off.
\begin{figure}[t]
\begin{center}
   \epsfig{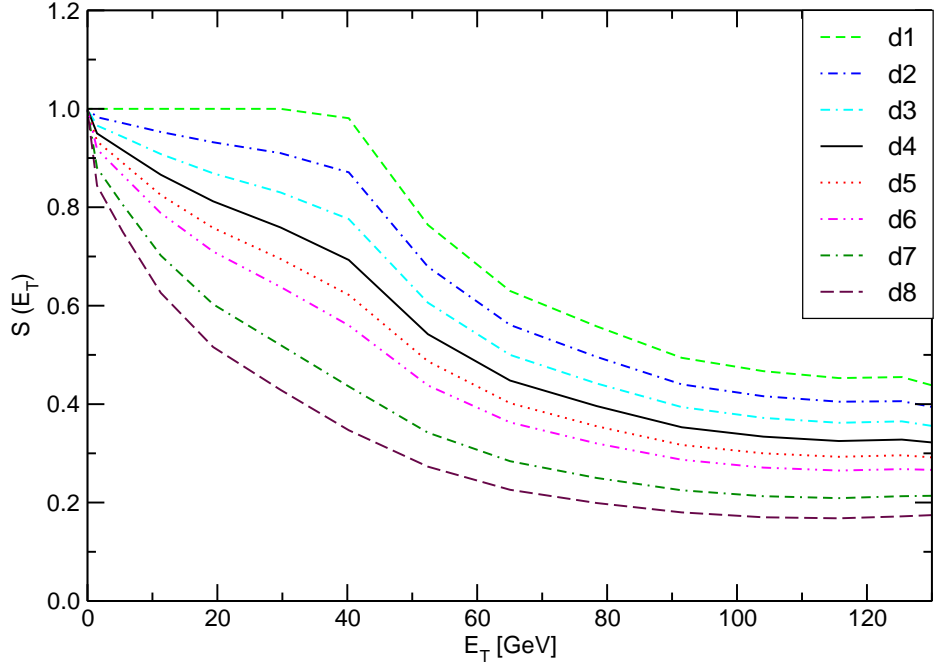}
\end{center}
\caption{The survival probability S($E_T$)
 in Pb--Pb collisions at the SPS
for the case of vanishing nuclear absorption, Matsui -- Satz mechanism
and $J/\psi$ dissociation of gluons present. In all cases $n_{g/nn}$ = 0.58,
(d1) $\sigma_{g\psi}$ = 0.0 mb; (d2) $\sigma_{g\psi}$ = 0.2 mb;
(d3) $\sigma_{g\psi}$ = 0.4 mb; (d4) $\sigma_{g\psi}$ = 0.6 mb;
(d5) $\sigma_{g\psi}$ = 0.8 mb; (d6) $\sigma_{g\psi}$ = 1.0 mb;
(d7) $\sigma_{g\psi}$ = 1.5 mb.
Other parameters fixed: $\epsilon_g$ = 0.8 GeV,
$\tau_{em}$ = 1 fm/c, $\epsilon_{c}$ = 0.7 GeV/fm$^3$.}
\label{fig3}
\end{figure}

In Fig.3 we see again that with increasing $\sigma_{g\psi}$ the survival
probability decreases, but the abrupt onset of increasing $J/\psi$
suppression is gradually washed out.

Finally, in Fig. 4 we present the results of the same calculation as
shown in Fig. 3 with only one difference: nuclear absorption with
$\sigma_a$ = 4.2 mb, \cite{NA50Nantes} is included.
\begin{figure}[t]
\begin{center}
   \epsfig{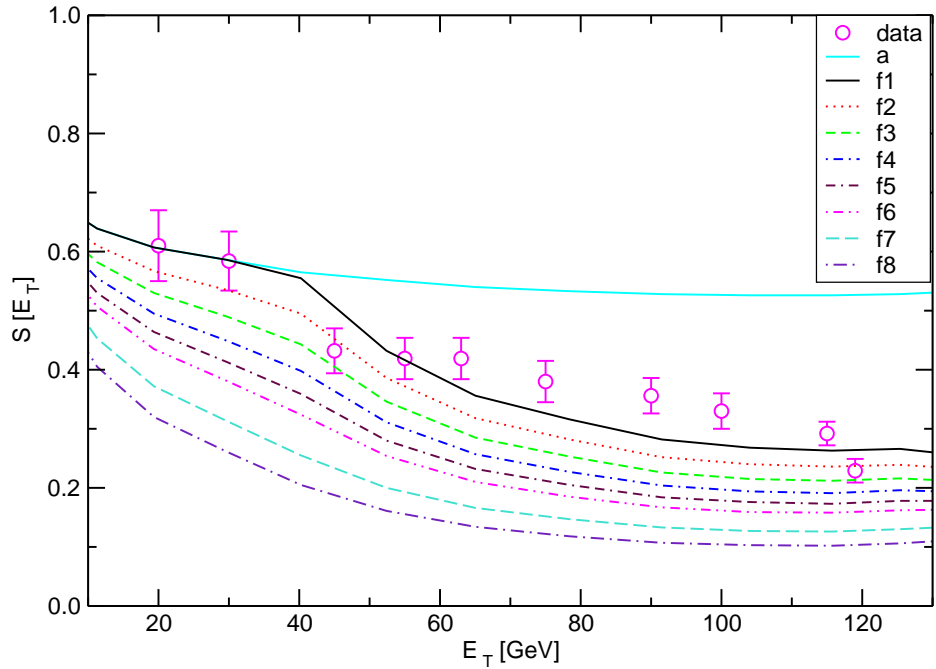}
\end{center}
\caption{The survival probability S($E_T$) in Pb--Pb
collisions at the SPS
for the case of nuclear absorption with
 $\sigma_a$ = 4.2 mb, Matsui -- Satz mechanism
and $J/\psi$ dissociation of gluons present. In all cases $n_{g/nn}$ = 0.58,
(a) only nuclear absorption. In further cases Matsui -- Satz mechanism included.
(f1) $\sigma_{g\psi}$ = 0.0 mb; (f2) $\sigma_{g\psi}$ = 0.2 mb;
(f3) $\sigma_{g\psi}$ = 0.4 mb; (f4) $\sigma_{g\psi}$ = 0.6 mb;
(f5) $\sigma_{g\psi}$ = 0.8 mb; (f6) $\sigma_{g\psi}$ = 1.0 mb;
(f7) $\sigma_{g\psi}$ = 1.5 mb.
Other parameters fixed: $\epsilon_g$ = 0.8 GeV,
$\tau_{em}$ = 1 fm/c, $\epsilon_{c}$ = 0.7 GeV/fm$^3$. Data
taken from Ref.\cite{QM02}}
\label{fig4}
\end{figure}
Results presented in Fig. 4 summarize the situation. The curve (a)
shows the well known fact that the nuclear absorption alone cannot
reproduce J/$\psi$ suppression in Pb--Pb interactions as observed
by the NA50 Collaboration \cite{QM02}. The curve (f1) agrees in
a qualitative way with the patterns of the data \cite{QM02}. This curve
corresponds to the Matsui--Satz mechanism \cite{MS86} in the KLNS
formalism \cite{KLNS97}. This is seen from from the calculation
of $n(t_1)$ as given by Eq. (12) and of the energy density of the
system of gluons according to Eq. (14). Note that for the curve
(f1) the cross--section $\sigma_{g\psi}$ for the dissociation of
$J/\psi$ by gluons vanishes. The curves (f2) to (f8) contain both
the Matsui--Satz mechanism and the dissociation of $J/\psi$ by
gluons, but $\sigma_{g\psi}$ is gradually increasing. The abrupt
onset of $J/\psi$ suppression at $E_T \approx $ 40 GeV is gradually
washed - out. This is also natural since with increasing $\sigma_{g\psi}$
more and more of $J/\psi$'s are suppressed by interactions with gluons
and the Matsui--Satz mechanism dissolves only those $J/\psi$'s that
survived interaction with gluons.

\section {Comments and conclusions}
\label {comments}
The purpose of the present paper was not to make
detailed analysis of $J/\psi$ suppression by the ISG.
 We were more interested in qualitative features of the
ISG model and its possibilities to approach
 the results following from models based on
$J/\psi$ suppression
 by QGP. More detailed analysis would certainly require
using Woods-Saxon densities, using more accurate relationship
between $E_T$ and $b$, respecting fluctuations, etc.
 But we are
convinced that qualitative features of
prediction of our version of the ISG model
and in particular of its differences with respect
to the QGP model
 would survive the improvements brought in by more
 sophisticated analysis.
The question of whether $J/\psi$ is suppressed by
QGP or by another
 mechanism will be finally decided by experiment, hopefully by
the forthcoming data by NA60 Collaboration. If the
anomalous $J/\psi$ suppression
 will turn up to be really as abrupt as possible, and this is the
case of QGP model, other interpretation will be in problems.
 If it will turn out to be less abrupt, it will be easy for
other models to fit the data.

As a by--product the present model provides a sketch of a possible
space--time picture of the dynamics of nuclear collisions at
the CERN SPS. The picture contains the following ingredients
\begin{itemize}
\item When nuclei traverse each other, each nucleon looses energy
in semi--hard nucleon--nucleon collisions. The cross--section of
this process is about 10mb and in each of the collisions two gluons
with energy of about 0.6 -- 1.0 GeV are produced. When traversing
the Lorentz contracted
length 2L$_B \approx $ 1.1 fm in the other nucleus, a nucleon
makes about 1.4 collisions and looses in average about 1 GeV of
energy.
\item During the time when nuclei cross each other there
is no production of hadrons or gluons with lower transverse
momenta because of reasons described in the Appendix,
see also Ref. \cite{LPPT}.
\item The fragmentation of nucleons starts only after the two tubes
have crossed each other.
\end{itemize}

The picture indicates why on the one hand the Wounded Nucleon
Model \cite{WNM} is a reasonable first approximation to particle
production and on the other hand it shows that production of
energy densities permitting the formation of QGP can be caused
by semi--hard processes.

\medskip
{\large \bf Appendix}

In nuclear interactions soft exchanges in nucleon- nucleon
subcollisions are
suppressed by the Landau-Pomeranchuk-Migdal (LPM) mechanism,
 for a review
see Ref.\cite{LPM}. If time interval and longitudinal
distance between two subsequent collisions of a nucleon
from nucleus A with
nucleons in nuclei B are $\Delta t$ and $\Delta z$ the
 LPM mechanism
suppresses processes with energy and momentum transfer obeying the
condition
\begin{equation}
\Delta p_z < \frac{\hbar}{\Delta z} ,
\quad \Delta E <\frac{\hbar}{\Delta t}
\label{eq26}
\end{equation}
Mean free path for a nucleon passing through a nucleus
at rest is about 3fm. When considering a collision of two nuclei
 at $E_{lab}\approx 200 GeV$, in
the c.m.s. of nucleon-nucleon collision both nuclei
are contracted by the
Lorentz factor $\gamma \approx 10$, the mean free path becomes
 shorter by the
factor $\gamma$ and Eq.(26) leads to suppression of
energy and momentum
transfers with
\begin{equation}
\Delta p_z < 600 MeV/c ,\quad  \Delta E < 600MeV
\label{eq27}
\end{equation}
The problem with estimating the density of intermediate
stage gluons in
nuclear collisions is due to the  fact that  there is no
reliable theory of
nuclear interactions in the CERN SPS energy region.
In this situation
 we can
use only models which might reflect to some extent what is really
happening.
The venerable Wounded Nucleon Model (WNM) \cite{WNM}
describes the
production of secondary hadrons in nuclear collisions
 as a fragmentation
of participating (wounded) nucleons. The fragmentation
 is independednt of
the number of preceding collisions. The model does not
describe details
of the transverse energy production in nuclear collisions and
proton stopping. Apparently there exists also some
 contribution to the
total transverse energy in nuclear collisions given
roughly by the number
of nucleon-nucleon interactions.

There are two groups of models indicating that even in the CERN SPS
energy region this does happen and partonic degrees
of freedom may play an important role.

First, in Glauber type models of multiparticle production, see
 Refs.\cite{DGS85,Lich1,Wong,Liet1} aiming to describe the proton
 stopping and the production of total transverse energy
 in pA (and AB) collisions one has to assume
 some loss of proton momentum in each of $\nu$ sub-collisions. For
 instance in Ref.\cite{Liet1} the author requires the degradation of
  proton's
 momentum by $\Delta y\approx 0.5$ (in the c.m.s. of proton-nucleon
 collisions). This means that the proton's momentum is changing in an
 SPS experiment  from the original
 10GeV/c (in the c.m.s.) to 6.05 GeV/c after
 the first collision, to 3.62 GeV/c after the second one, to 2.13 GeV/c
 after third one, and to 1.25GeV/c after the fourth one.
  The corresponding momentum losses are: 4 GeV/c; 2.4 GeV/c;
 1.3 GeV/c and 0.8 GeV/c. The average of these four momentum losses is
 2.13 GeV/c, and although this indicates that
  perhaps the collisions  may be considered as semi-hard and that
 the partonic degrees of freedom might be relevant.

In models of this type there has always been a problem of why
secondary hadrons assumed to be produced in individual nucleon-
nucleon collisions do not cause intense cascades in the region
of $E_{lab}$ of a few hundreds of GeV. The standard solution of
the problem is to introduce for secondary hadrons a formation
time $\tau_f$ with Lorentz dilated
$\tau_f$ responsible for the attenuation of cascade effects.

 Second, in their analysis of the magnitude of the nucleon- nucleon
 non-difractive cross- section  ($\sigma_{ND}$) the authors of
  Refs.\cite{AGM,SvZ,DuPi}  have shown that most of or the whole
   of $\sigma_{ND}$ can
   be described as due to semihard QCD processes.
Abou-El-Naga, Geiger and
M\"{u}ller \cite{AGM} have calculated
$\sigma_{ND}=\sigma_{QCD}$ by using
 the perturbative QCD expressions (in the lowest order)
 for the elastic parton-parton scattering. The parameter
 $Q^2$ entering the structure functions and the $\theta$- function
  is taken as $Q^2=p_T^2=\hat u\hat t/\hat s$ with $\hat s=x_1x_2s$.
 Cross-sections for $Q^2\le (p_T^{min})^2$ were put equal to zero.
 The value of $p_T^{min}$ was taken as a free parameter
to be determined
 by the requirement that the whole of $\sigma_{ND}$ is given by the
 perturbative QCD calculation.
   Structure
  functions were taken either from Eichten et al.\cite{Eich} (EHLQ) or from
Gl\"{u}ck et al.\cite{GRV}.

 The magnitude of $\sigma_{ND}$ depends in a crucial way
on the value of
$Q^2$ at a given value of s. The measured value of
$\sigma_{ND}$ is equal
to $\sigma_{QCD}$ (most of that being due to gg interaction)
 provided that
$$
\sqrt{Q^2}=p_T^{min}=p_0\left(\frac {s}{GeV^2}\right)^{\alpha}
$$
with parameters $\alpha=0.115$ and $p_0=)0.485GeV/c$ for the EHLQ
structure functions and $\alpha=0.151$ and $p_0=0.357$ for
GRV structure
functions.

For s=400 GeV${}^2$ corresponding to the CERN SPS this gives
 $p_T^{min}$=
0.97 GeV/c for the EHLQ case and 0.88 GeV/c for the GRV one. Of course,
the pQCD is used in the situation when
extrapolations are not reliable,
neither in what concerns the $Q^2$ dependence
of structure functions nor
in calculations of parton- parton cross- sections. But it is still
reasonable to assume that some fraction of
$\sigma_{ND}$ is due to parton- parton (mostly gg) interactions.

\medskip
{\bf Acknowledgements}

One of the authors (J.P.) is indebted to the CERN Theory Division for the
hospitality extended to him. The work of J.P. and N.P. has been supported
also by the grant of the Slovak Ministry of Education VEGA F2V13.

\end{document}